\documentstyle[epsfig]{article}
\def\parn{\par\noindent}
\def\be{\begin{equation}}
\def\ee{\end{equation}}

\begin{document}
\title{Hadronic Masses and Regge Trajectories}
\author{Silvana Filipponi\footnote{Silvana@pg.infn.it}[1,2] 
and Yogendra Srivastava\footnote{Srivastava@pg.infn.it} [2,3]}
\maketitle
\centerline{HUTP-97/A093} \vskip5mm
\centerline{1.{\it Physics Department, Harvard University, Cambridge MASS, 
USA.}}
\centerline{2.{\it Dipartimento di Fisica e INFN, I-06100 Perugia Italy.}}
\centerline{3.{\it Physics Department, Northeastern University, Boston MASS,
USA.}}
\begin{abstract}
A comprehensive phenomenological analysis of experimental data and some 
theoretical models is presented here (for mesons) to critically discuss how 
Regge trajectory parameters depend on flavor. Through  
analytic continuation of physical trajectories (obtained from resonance data) 
into the space like region, we derive the suppression factor for heavy
flavor production. The case of our D Regge exchange, both for D and 
$\Lambda_c$  production, is considered in some detail. Good agreement with 
data is reached confirming that indeed the slopes of heavier flavors 
decrease. This result suggests that the confinement potential has a 
substantial dependence on the quark masses. In a simple non-relativistic 
model, constrained to produce linear Regge trajectories, it is shown
that a linear quark mass dependence is required (in the confinement part  
of the potential) in order for the slope to decrease in the appropriate way. 
\end{abstract}
\section{Introduction}
Hadron internal dynamics is described by QCD but in deriving their
properties many problems arise essentially due to the mathematical 
complexity of  relativistic bound state problems in 
non abelian gauge  theories. This has led to techniques and approximation 
schemes  
to perform phenomenological calculations which differ according to the 
species of the hadron. Since QCD is flavor independent but the hadrons 
are not, it is only natural that quark masses which distinguish flavor 
appear as crucial quantum numbers not only in fixing the scale of the
hadronic masses but in determining the choice of the approximation itself.
\par
A relativistic or non-relativistic model is usually invoked 
depending 
upon whether the ratio ${{m}\over{\Lambda_{\chi}}}\ < \  
1$ or to $> \ 1$, where $\Lambda_{\chi}$ is the hadronic scale  
and $m$ the quark mass.  
It would then appear that the dynamics for light quarks should be largely 
detemined by $\Lambda$ and by $m$ for heavy flavors. The case of heavy- light  
systems would lie somewhere in between. The purpose of this work is to 
investigate the question of quark mass dependence in a systematic way.\par 
Our analysis is based on the construction of Regge trajectories under the 
the following two hypotheses. First, we assume all trajectories to be 
linear in the squared mass (of the hadronic state) for all flavours. 
\be
\alpha(s) =\alpha(0)+ s\: \alpha '
\label{regge}
\ee
For light baryons and mesons, we have ample and compelling phenomenological 
evidence supporting this assumption. Our analysis finds no evidence for
strong deviations from linearity also for heavier flavors.\par
The second assumption is that the functional dependence of the
two parameters $\alpha(0)$ and $\alpha'$ on quark masses is through the
combination ($m_1 +\: m_2$). Arguments in support of this hypothesis are 
presented in {\bf section two}.\par
In {\bf section three} through an analysis of Regge 
trajectories for mesons of all flavors, we  obtain  analytic
forms for the slope as well as the intercept parameters as a function of 
($m_1+m_2$). The ``distances'' between trajectories are also investigated 
in the same fashion. In {\bf section four} trajectories are constructed 
on which physical mesons lie. In {\bf section five}, we employ their analytic 
continuation into the space like region to discuss how heavy flavour 
production is suppressed. The case of D meson trajectory exchange
is analysed in some detail. Our model predictions are compared successfully
with available experimental data. In {\bf section six} we address the
implications which our Regge trajectories may have on the dynamics underneath.
The decrease in the slope for heavier flavors is argued to imply that the 
confining potential is strongly flavor dependent. A non-relativistic 
potential model exhibiting linear trajectories is constructed where a 
simple linear dependence on the quark mass (for the confinement term) 
reproduces the appropriate behavior in the slope.
\section{Choice of the variable ($m_1+m_2$)}
Let us consider a meson of mass $M(m_1, m_2)$ as a bound state of a quark of 
mass $m_1$ and an anti-quark of mass $m_2$. Quite generally, M may be  
parametrized as a function of $\rho=\sqrt{m_1m_2}$ and 
$\tilde{m} = (m_1+m_2)$. 
Since quarks have never been observed as free particles, there is no unique  
definition of $m_1$ and $m_2$. Transcending the bound state dynamics,
this uncertainty further complicates how $M$ depends on $\rho$ and 
$\tilde{m}$. For the constituent quark picture adopted here, we present 
arguments below in favor of $\tilde{m}$ being the relevant parameter upon 
which $M$ depends, for practically all cases where data are available. 
In some cases, where the dependence on  $\rho$ turns out to be 
small but not negligible, 
it may be incorporated perturbatively. \par
 As a widely accepted convention, quarks are considered light if 
$m<\Lambda_{\chi}$ (u,d,s) and heavy if $m>\Lambda_{\chi}$ (c,b,t), 
indicating that very different properties arise depending on this scale 
($\Lambda_{\chi} \approx 1$ GeV). Given the above classification for 
quark masses, there are four types of mesons:
(i) both quarks are light; (ii) both quarks have the same mass; 
(iii) one quark is light and the other heavy; (iv) both quarks are heavy.
For case (i), both $\rho$ and $\tilde{m}$ are
much smaller than  $\Lambda_\chi$, hence negligible in our scale for 
all quark masses. For case(ii),
$\rho$ and $\tilde{m}$ degenerate into a single parameter. For case (iii),
$\tilde{m}$ is the only remaining parameter. Hence, the question 
whether $\tilde{m}$ is the dominant parameter 
is of relevance only for case (iv). For the subclass of (iv) which can be 
described via non-relativistic dynamics, the dependence on $\rho$ (or, 
equivalently on the reduced mass) must by implication be small. In such 
cases, it can be included perturbatively. For the rest of class (iv), 
the dependence on $\rho$ cannot a priori be guaranteed to be small. In
practical terms, this may be quite relevant only for the ($b\bar{c}$) system,
where appreciable differences in masses may occur in parametrizations
with or without the inclusion of terms containing $\rho$. Thus, data on 
this system would be crucial in distinguishing between the two choices. 
\par
All the analysis of the next section will be performed using the $\tilde{m}$ 
variable. Without the above justification, the same choice has also been 
made in \cite{IMZ91}. Corrections due to electric charge or isospin are not
included here. 
\section{The phenomenological analysis}
\subsection{The $\alpha '$ and $\alpha _I(0)$ parameters of the 
Regge trajectories}
In this section, we derive analytic forms for  the  slope and intercept 
parameters for the Regge trajectories for all mesons. For this purpose, 
we use experimental data \cite{DP96} as well as results from theoretical 
models. Theoretical information is required to supplement the experimental 
one for mesons composed of heavy quarks where only the lowest part of the 
spectra has been measured. We selected some models which 
fit the measured energy levels and predict the masses for  higher resonances. 
For some states, quark model needs to be invoked to fix the spin
\cite{DP96}, whenever it has not been determined experimentally.  
The spin quantum number being crucial for our analysis, it is 
essential to verify whether  the quark model prediction is confirmed
by other independent calculations. As experimental spectra are almost 
unkown for the B mesons and poorly measured 
for the D mesons, we do not attempt for them the same detailed analysis 
as has been done for charmonium and bottomium. For the D system, only a rough 
determination of the Regge trajectories can be made from experimental data. 
\par 
While using experimental data, our choice for the quark masses are 
as follows:
\be 
 m_u=m_d=0.03\:GeV,\ \  m_s=0.5\:GeV,\ \  m_c=1.7\:GeV \ \ and\ \  
 m_b=5\:GeV .
 \ee
While using inputs from theoretical models, values for quark masses 
are fixed according to the ones used in the particular model. \par 
\vspace{0.1cm}\par
Let us begin with the slope parameter. In Fig.1 input data for the slopes 
are presented. \par
For light mesons, we refer to a recent analysis  of  Regge trajectories 
in \cite{T90}, and to the MIT bag model\cite{JT76} which gives the 
theoretical value for $\alpha'_{bag}\approx 0.9\:GeV^{-2}$. In this model, 
the bag is supposed to be a rapidly rotating, linearly extended object, 
where the massless quarks are fixed at the  ends; the slope parameter  is 
obtained through a classical calculation of the energy and angular 
momentum of the bag. \par 
  The four experimental points 
(the bullets) for light mesons, have been determined analyzing leading Regge 
trajectories for systems with different isospin (along  with their  
quark content): \parn
I=0,  both $\omega$ ($u\bar u+d\bar d$ bound states) 
and $\phi$ ($s\bar s$ bound states) resonances- with slightly different 
slopes;\parn 
I=1, $\rho$ resonances;\parn 
I=1/2, K resonances.\parn 
For charmonium, we calculate the slope parameter using the masses of the
two states that have been experimentally measured both for the leading  
[$J/\psi(1s)-\chi_{c2}(1p)$] and for the second trajectory [$\eta_c(1s)-
\chi_{c1}(1p)$] which differ by one unit of spin. A similar situation is
found  for bottomium. For both these systems  a more complete spectrum is  
provided by theoretical models (\cite{R79},\cite{M80},\cite{GRR93}). We check
that an almost linear behaviour is reproduced with the predicted masses for  
the higher J states. For both systems, the calculated slopes are very close
to the ones derived from our analysis. This is not an obvious result because  
the model predictions for the unkown J states are given in a completely  
independent way without imposing any linear behaviour.    
The theoretical models we analyse  are non relativistic potential  models  
(relativistic correction terms are included by hand) where  a central potential  
describes the interaction. In \cite{R79} a smooth function interpolating  
between the Coulomb and the linear term fits the spectra. The other choices 
for the potential are: $V(r)=A+B\;r^{0.1}$ for \cite{M80} and 
$V(r)=\tilde{A}+\frac{\tilde{B}}{r^{0.1}}$ for \cite{GRR93}. 
For these three models, the slopes are calculated and included in  our 
analysis.\par
Experimental data for the D and $D_s$ mesons  are used to obtain their 
slope.  
Predictions  from a semirelativistic model in \cite{MKM88} for the D and
B mesons are also included.\par 
All the inputs described above have been used in our analysis and are shown
in  Fig.1, along with a global fit. The analytic behaviour for the slope is
found to be 
\be
\alpha '=\frac{0.9}{1+0.2\:(\frac{m_1+m_2}{GeV})^{3/2}}\: GeV^{-2}
\label{apt}
\ee
\indent 
An identical analysis has also been carried out for $\alpha _I(0)$, where the  
subscript I refers to the leading Regge trajectory. All input data points  
are shown in Fig.2 along with the following analytic  expression
\be
\alpha _I (0)=0.57-\frac{(m_1+m_2)}{GeV}
\label{at}
\ee
In Fig.2 only two entries (they refer to the $B_c$ system)   
have too small a value compared to the general trend. 
We are unable to explain these deviations from the expected behaviour. 
\vskip5mm
Before concluding this section, we consider some consequences following 
from our result for the slope parameter, eq.\ref{apt}. 
The 3/2 power implies a rather large value for the kinetic energy possessed
by bound state of heavy quarks.  
This becomes substantial already for the $B_c$ system, and is certainly 
so for toponium and top mesons. Unfortunately, mass spectra are
completely unkown in the first case \cite{DA97}, 
while top bound states are predicted neither to form,
due to the large mass and short life time (through weak decays)   
of the top quark.
Presently, it is  not possible to verify from data 
whether such a large kinetic energy term  exists.\par 
It must be emphasized that parametrizations which include other quark 
mass dependences than $m_1+m_2$ alone, might change significantly the 3/2 
power law. The accuracy of present data renders difficult such an analysis 
in terms of more than one combination of the quark masses. \par
In string type models \cite{KB}, where the associated kinetic energies are 
not overly large, $1/\alpha '$ would be linear in $\tilde{m}=m_1+m_2$
(within our assumption that $\alpha '$ only depends upon $\tilde{m}$). 
Hence, we tried to constrain the fit to reproduce
this behavior. A phenomenological problem arises however with the 
light sector. Since, $\alpha '=\frac{\alpha '(0)}{1+ A\:\tilde{m}}$  
has a large negative derivative for  small $\tilde{m}$, there appear 
significant variations for the slope in that region. On the
other hand, we require for all the slopes of the light sector-
$\alpha '_{light}\approx (0.8\div 0.9)\:GeV^{-2}$. Such can be obtained  
only with an almost perfect degeneracy in the u,d and s quark masses.\par  
We hope to return to this interesting but delicate  question elsewhere.
Below we continue our analysis with the unconstrained fit, 
eq.\ref{apt}, which reproduces the observed fall off in the slope quite well. 
\subsection{Level splitting through the 
$\alpha (0)\equiv\alpha _I(0)-\alpha_{II} (0)$ parameter.}
Let us now consider the spacing between the first and successive    
trajectories for various mesons. This distance is defined to be the energy 
squared gap between states of fixed J for successive trajectories,   
$[E^{I+1}_{J}]^2-[E^{I}_{J}]^2$. We  also analyse 
the corresponding energy level splitting $[E^{I+1}_{J}]-[E^{I}_{J}]$.  
\par 
There is phenomenological evidence 
suggesting that successive trajectories alternate between states of
normal and abnormal parity. That is, the 
first trajectory contains the normal set starting with $J^P=1^-$, 
followed by the abnormal set with $J^P=0^-$, and so on. In particular,  
lowest J=1 states for any meson \cite{DP96} has negative parity. As energy 
increases, we encounter a state of positive parity, followed by a negative 
parity  state and so on. This trend is summerized in Table I.
\begin{center}
\begin{tabular}{|l|l||l|l||l|l||l|l||}
\hline
\multicolumn{8}{|c|}{ {\bf J=1 states}}\\ \hline \hline
State        &P&State&P &State&P &State&P  \\ \hline \hline
$\omega(782)$&-&$h_1(1170)$&+&$\omega(1420)$ and $\omega(1600)$ &- &?& +\\
 $\rho(770)$ &-&$b_1(1235)$&+&$\rho(1450)$&- & ?& + \\
$K^*(892)$   &-&$K_1(1270)$&+&$K^*(1410)$ & -& $K_1(1650)$& + \\
 $\phi(1020)$&-&$h_1(1382)$&+&$\phi(1680)$&- &? & + \\ \hline
$D^*$&-&$D_1(2420)$&+&? &- &? &+ \\ 
$D_s^*$& -&$D_{1s}(2536)$&+&? & -&?& +\\ \hline
$J/\psi(1s)$ &-&$\chi_{c1}(1p)$&+&$\psi(2s)$ and $\psi(3770)$&-& ?& + \\ \hline
$Y(1s)$      &-&$\chi_{b1}(1p)$&+&$Y(2s)$&-&$\chi_{b1}(2p)$ &+\\ \hline
\end{tabular}
\vskip3mm
Table I: {\em  Spectrum of spin one states.}
\end{center}
\vskip5mm \noindent
Exceptions to this regularity are present  and we  discuss them briefly 
in the following. For example, for the omega resonances, we have two
closely lying states, $\omega (1420)$
and the $\omega (1600)$, which have both been assigned  
negative parity.  A similar situation is found for the charmonium states, 
$\psi(2s; 3686.00\pm0.09 MeV)$ and $\psi(3770)$  states. \par
Another possible exception concerns the
$\hat {\rho}(1405)$, $J^P=1^-$  state, which was observed but not 
reconfirmed. In fact this state is omitted from summary table, even if 
reported in \cite{DP96}. If such a state did exist, it would be out 
 of the scheme presented above.\par
On the other hand, for K mesons the situation is extremely regular. 
First three states follow this pattern. Then somewhat doubtful 
K(1650) of positive parity is followed by  $K^*(1680)$ of negative parity.
This last entry is not shown in the previous table for lack of space.\par
As can be seen in Table I, data are scarce for D mesons. In any case, the 
known states follow the regular pattern and we have included them in our 
analysis. For the first entry on line 6 of Table I, the state labelled 
$D_s^*$, quantum numbers have not yet been measured. The assignment 
$J^P=1^-$ for this state appears consistent with the general pattern.
Hence, we have included it in the first column.\par
 Finally for the charmonium and 
bottomium systems, beyond the tabulated  states, a series of unconfirmed 
P=- states have been reported [$\psi(3770)$, $\psi(4040)$, $\psi(4160)$, 
$\psi(4415)$ for charmonium and Y(4s), Y(10860), Y(11020) for 
bottomium]. Apart from these possible exceptions, the alternating behaviour 
between normal and abnormal sets of Regge trajectories 
seem to be  confirmed.\par 
In  Fig.3, energy splitting and the energy squared gap  
between spin one states  from columns I and II of Table I
are plotted as a function of $m_1+m_2$.  
No input from theoretical models are available  since   only 
normal states were considered there. \par
As a general behaviour, in Fig.3 
energy splitting seems to be almost 
constant around 0.4 GeV for all existing mesons, while energy
squared increases. \parn
The distance between the normal  and the successive abnormal trajectory 
is defined through the energy squared gap for a fixed J
\be
[\bar{E}_{J}]^2-[E^{I}_{J}]^2=\frac{1}{\alpha '}\: 
\left [ \alpha_{I}(0)-\bar{\alpha}(0) \right ].
\label{distance}
\ee
The energy splitting is given by 
\be
\bar{\Delta} E_{J}\equiv \bar{E}_{J}-E^{I}_{J}=
\frac{1}{\sqrt{\alpha '}}
\;\;\left\{\sqrt{J-\bar{\alpha} (0)}-\sqrt{J-\alpha _{I}(0)}\right \}
\label{distances}
\ee
where the tilde superscript refers to the abnormal and the
index I to the leading trajectory. 
From Fig.3, we find that $\bar{d}\alpha (0)\equiv 
\alpha_{I}(0)-\bar{\alpha}(0)$ depends weakly upon $m_1+m_2$
and this dependence has been ignored. 
A conservative estimate for this parameter is given by
\be
0.6<\bar{d}\alpha (0)\equiv \alpha_{I}(0)-\bar{\alpha}(0)<1.1
\label{a1-a2}
\ee
For $\alpha '$ and $\alpha_{I}(0)$ we use our analytic results given in
eq.\ref{apt} and eq.\ref{at}  to obtain the continuous curves 
shown in Fig.3.
We repeat the same analysis  for the distance between the 
leading and the second trajectory, using both the experimental data  
(columns I and III from Table I) and theoretical estimates from models  
discussed before. In Fig.4, we plot the energy squared  gap and, in Fig.5, 
the energy gap between the leading and the second trajectory, as a function
of $m_1+m_2$.  From these curves, we deduce
\be
d\alpha (0)\equiv \alpha _{I}(0)-\alpha _{II}(0) \approx (1.3\div 1.6)
\ee
These data, even though not very precise, seem to indicate that
the separations between the leading normal and the successive abnormal and
between the abnormal and the second trajectories are not equal.
\par
\section{Leading trajectories for heavy mesons}
In section three we derived how Regge parameters 
functionally depend on $m_1+m_2$. 
As a function of the sum of the constituent quark masses and the 
rest mass of the corresponding bound state, the leading trajectory is 
found to be
\be
J\:(m_1+m_2,\:M^2)=0.57-\frac{(m_1+m_2)}{GeV}+
\frac{0.9\:GeV^{-2}}{1+0.2\:(\frac{m_1+m_2}{GeV})^{3/2}}\;M^2
\label{allregge}
\ee
For $m_1+m_2\approx 10$ GeV the slope
of the corresponding  trajectory (describing bottomium) is close to
zero. \par 
The effective $m_1+m_2$ parameter for any meson is derived inserting 
into  eq.\ref{allregge}  the experimental value for the mass of the lowest 
energy, J=1 state.
The precision of this analysis is estimated to be of order of 
ten percent. At the present level, eq.\ref{allregge} cannot be used to 
``deduce'' the quark masses.  \par
About the $D_s$ and $B_s$ mesons, the lowest energy J=1 state has not
yet been detected, \cite{DP96}. As possible candidates,  
we choose $D_s^*=(2112.4\pm 0.7)\:MeV$ (the same choice has been made in 
section three) and $B^*_s=(5416.3\pm 3.3)\:MeV$.  
These states are used here to derive  the slope and intercept 
parameters for $D_s$ and $B_s$ systems. \par 
This tecnique has been used to calculate rest mass spectra. The ``mass
formula'', as well as results for D and B mesons, 
charmonium and bottomium spectra  are presented in another paper \cite{FS97}. 
In  Table II, the slope and intercept parameters have been calculated for 
all the heavy mesonic sector, except for the $B_c$ system, whose ground
state has not yet been detected  \cite{DA97}, and toponium and top mesons 
(see \cite{FS97} for a detailed discussion about mesons composed of  the top
quark and their Regge trajectories).
\vskip2mm 
\begin{center}
\begin{tabular}{|l||ll|}
\hline
Meson&$\alpha _I(0)$&$\alpha '$ ($GeV^{-2}$)\\ \hline \hline
D&-1.35& 0.59 \\
$D_s$& -1.51&0.56\\
Charmonium&-2.83&0.40 \\
B&-5.45&0.23\\
$B_s$&-5.55&0.22\\
Bottomium&-9.67&0.12\\ \hline
\end{tabular}
\vskip3mm
Table II:{\em Our results for the slope and intercept parameters of 
heavy mesons.}
\end{center}
\vskip2mm
About the slope parameter, we find a very good agreement with independent 
calculation in \cite{IMZ91}, which gives  $1/\alpha '=2.48\;GeV^2$ for 
charmonium and $1/\alpha '=6.21\;GeV^2$ for bottomium.\par
In Fig.6  our leading trajectories for all existing flavours are plotted 
to show  the intersection region.
About the light sector, that has not been analysed in this work, only one 
trajectory is drawn for all  isospin and the corresponding  
parameters are calculated for  $m_1+m_2=0$.
\section{Predictions for the space-like region}
Up to now our analysis has been concerned with physical states lying on the
trajectories. Of course, Regge theory allows for
applications in the space-like region through 
predictions for hadronic scattering processes, since its asymptotic behaviour 
in energy is related to the exchanged Regge trajectory.   
Exchanges of light trajectories for exclusive processes 
have been quite successful in the past. 
For the heavy sector, only inclusive data are available for 
charm  production.  
We employ the analytic continuation  of the D Regge trajectory to  compute 
the inclusive production of D mesons and $\Lambda_c$ baryon  
through the di-triple Regge formalism. These results are compared with  
available experimental data.\par   
Such an  application  to the  space-like region 
is a completely independent check for our constructed Regge trajectories 
since other data and a different part of the phase space  are now  involved.
Similar data are not available for the B system, thus only theoretical 
predictions are given. 
\subsection{The suppression factor for  heavy flavour production}
For an inclusive reaction $a+b\rightarrow c+X$, Regge theory predictions
are given in the Tri-Regge asymptotic limit, \cite{M70}, where
$t=(p_a-p_c)^2$ is kept fixed, 
both the c.m. energy $s=(p_a+p_b)^2$ and $M_X^2=(p_a+p_b-p_c)^2$  are large, 
but $M_X^2/s$ is small.
For an inclusive process the differential cross section is given by, 
\cite{PS71}
\be
\frac{d^2\sigma}{dM_X^2\:dt}\approx \frac{\gamma_0(t)}{s}\;\left (
\frac{M_X^2}{s}\right ) ^{1-2\alpha (t)}
\label{regsca}
\ee
where $\alpha (t)$ is the Regge trajectory for the exchanged particle.
Momentarily neglecting the t dependence, we have
\be
\frac{d\sigma}{dM_X^2/s}\approx 
\left (\frac{s}{M_X^2}\right )^{-n}\hspace{1cm} n=1-2\alpha(0)
\label{suppr}
\ee
showing the energy suppression factor in the differential cross section for 
inclusive heavy flavour production.  The power n  is related  to $\alpha(0)$,  
hence it linearly  increases  with the sum of the quark masses constituting
the 
meson. Thus, according to eq.\ref{at}, the suppression factor becomes larger 
as one goes from the D meson to bottomium. Predictions for all  heavy
flavor sectors are gived in Table III.
\vskip2mm 
\begin{center}
\begin{tabular}{|c|c|}
\hline
Meson& n\\ \hline\hline
D&3.7 \\
$D_s$&4.0\\
Charmonium&6.7 \\
B&12\\
$B_s$&12\\
Bottomium&20\\ \hline
\end{tabular}
\end{center}
\vskip3mm
Table III:{\em The power n controlling  the suppression term, 
eq.\ref{suppr}, appearing in the
differential cross section for heavy flavor production.}
\vskip2mm
\subsection{Inclusive charm production.}
The Fermilab experiment E769, \cite{A92}, has detected  charged and 
neutral D mesons 
 through  250 GeV $\pi ^-\; N$ interactions, using targets of Be, Al, 
Cu, W. There the differential cross section 
for charm meson production is analyzed using 
  the Feynman-x ($x_F$) and transverse momentum ($p_T^2$) variables.  
The  following fit has been  proposed in the  factorized form for 
the range $0.1<x_F<0.7$, $(0<p_T^2<4\:GeV^2)$ 
\be
\frac{d^2\sigma}{dx_F\:dp_T^2}\approx (1-x_F)^n \:e^{-bp_T^2}
\label{fitting}
\ee
The same  interaction  was investigated at CERN in two different experiments
(NA27,\cite{Ag85}  and NA32,\cite{B91}) using different targets, 
beam energy and   ranges for the $x_F$ and $p_T^2$ variables. Particulars
for all three experiments are shown in  Table IV below, along with the best
values for n and b. 
As we can
see all three independent measurements give  practically the same result for
both  parameters. \parn 
\vskip3mm\noindent
\begin{center}
\begin{tabular}{|l|l|l|l|}\hline
{\bf Expt.} & {\bf E769} & {\bf NA32} & {\bf NA27} \\ \hline \hline
$p_{beam} (GeV)$& 250 & 230 & 360 \\
Target & Be, Al, Cu, W & Cu & H \\ \hline \hline
$x_F$ fit range & 0.1 to 0.7 & 0.0 to 0.8 & 0.0 to 0.9 \\ \hline
n & $3.9 \pm 0.3$ & $3.74 \pm 0.23 \pm 0.37$ & $3.8 \pm 0.63$ \\ \hline \hline
$p_T^2$ fit range ($GeV^2$)& 0 to 4 & 0 to 10 & 0 to 4.5 \\ \hline
b($GeV^{-2}$)& $1.03\pm 0.06$ & $0.83 \pm 0.03 \pm 0.02$ & $1.18 \pm 0.18$ \\ 
\hline
\end{tabular} 
\end{center}
\vskip3mm
Table IV:{\em Values of n and b  for all three experiments.}
\vskip2mm\noindent
A similar value for n ($n=3.69^{+0.74}_{-0.71}$) 
is obtained  in \cite{B90} for  the 
  charmed baryon $\Lambda_c^{+}$ production from 230 GeV $\pi ^-$Cu and
$K^-$Cu interactions.\par
Both in $\pi^-\:N\rightarrow D\:X$ and  
$\pi^-\:N\rightarrow \Lambda_c\:X$,  a D trajectory is exchanged;   
Fig.7   represents this exchange for the $\pi \: N\rightarrow D\: X$ 
reaction.\par 
In order to test our predictions from Table III, we define the ratio 
between theoretical and experimental cross section as follows 
\be
R(M_X^2,t)=\frac{[d^2\sigma/dM_X^2\:dt]\:_{THEO}}{[d^2\sigma/dM_X^2\:dt]
\:_{EXP}},
\ee
where the numerator is given in eq.\ref{regsca}. For the denominator, 
eq.\ref{fitting} must be tranformed from the $(x_F,p_T^2)$  
into the $(M_X^2,t)$ variables and calculated in the Tri-Regge limit. 
The transformation along with some other kinematic 
details are given in the appendix, where we also show that  
$M_X^2/s\approx 0.3\div 0.6$ (or equivalently  $x_F\approx 0.4\div 0.7$)\
is the reliable region for testing our predictions. \par
Thus
in the triple-regge limit, for a fixed t value, the ratio R is given by 
\be
R\left( \frac{M_X^2}{s}\right)|_t \approx 
\frac{(1-\frac{M_X^2}{s})^{1-2\alpha(0)}}{(1-\frac{M_X^2}{s})^n}\;
\frac{\gamma_0(t)}{1-\frac{M_X^2}{s}}\;
\frac{(\frac{M_X^2}{s})^{-2\alpha'(0)t}}{e^{bm_D^2+b(t-m_D^2)
(1-\frac{M_X^2}{s})}}
\label{R}
\ee 
where we have neglected the t/s contribution from  eq.\ref{den} 
as explained in the appendix; $\alpha (0)$ refers to
the D trajectory. For small t, the first 
factor is dominant. For Table III, we find that our result for the power 
n is equal to 3.7   in  good agreement with the experimental  value  
from all  three experiments.  
To  show that the second factor in eq.\ref{R}  does not disturb
significantly the previous analysis, in Fig.8  we  plot $R/\gamma _0(t)$
for small and fixed  t values. As expected no large variations are found.   
\section{Possible dynamics for the phenomenological slope behaviour}
From our analysis it appears that the slope parameter depends  
on the constituent quark masses through a  3/2 power law. 
In the following we present some simple considerations which may lead to such 
a behaviour.\par 
It has already been shown \cite{NPS84}, \cite{DDD} that 
in a non relativistic approximation for the potential 
\be
V(r)=a\:r^p
\ee
p=2/3  generates linear Regge trajectories.  
In  such a  model,  the total energy is simply given by the 
centrifugal term and a confinement one, which is assumed to depend on the 
sum of the constituent masses ($\tilde{m}=m_1+m_2$) through an arbitrary 
function, $a=a(\tilde{m})$
\be
E(r;p)=\frac{J^2}{2\Lambda r^2}+
\frac{\Lambda (\Lambda r)^p}{p}\:a
\ee 
$\Lambda$ is a mass parameter and we require  the  centrifugal term to be 
$\tilde{m}$ independent in order to preserve a regular behaviour in the 
limit where the sum of the constituent masses vanish, i.e. the light mesonic 
sector. In such a  model, a quasi linear p=2/3 potential gives the 
phenomenological linear trajectories, when calculating the 
squared energy at the equilibrium point  $r_0$ 
\be
r_0=\frac{1}{\Lambda} \left (\frac{J^2}{a}\right )^{3/8}
\ee
Through a  simple calculation, one can derive for the slope of the 
trajectory  
\be
\alpha '(\tilde{m})=\frac{1}{2\Lambda ^2}\:\frac{1}{a^{3/2}}
\ee 
In order to obtain our phenomenological result, eq.\ref{apt}, 
$a(\tilde{m})$ must be proportional to $\tilde{m}$ suggesting 
that a linear mass term is required in the confinement 
potential in order to obtain the 3/2 power  in the Regge slope. 
\section{Conclusions}
In the present work, we have approached the generic problem of hadron
dynamics  
through considerations about the Regge trajectories for  mesons of all 
flavors. Previously, Regge phenomenology has been succesfully applied 
in the light sector for both  space- and time- like regions. The analysis 
of the Regge parameters for various flavors which we have 
performed does appear to justify the Regge technique as  a powerful 
one.\par 
As  has been shown in section five, heavy flavour production from
inclusive scattering is strongly suppressed with increasing energy.  
We succesfully checked our model predictions with available experimental data.
Work is still in progress to investigate further how the bound states 
depend on flavour. An important  task is to determine the 
relevant mass parameter describing the internal dynamics of the mesons. 
Until now, our choice of the sum of the
constituent quark masses has been  succesful. The  phenomenological
analysis does not show whether another combination of $m_1$, $m_2$ 
may also be relevant, yet, to this purpose, indirect applications of our 
results 
can be critically considered. Another interesting  question is how to generate 
the observed mass dependence. A plausible choice, which reproduced the 
observerd phenomenology, has been discussed in section six, where a mass
dependence was included in the confinement term.  \par
\vskip 5mm\noindent
{\bf Acknowledgements}\par\noindent
S.F. would like to thank  Professor S. Glashow
and the Harvard Physics Department for their hospitality.  

\vskip10mm
\begin{center}\Large{{\bf Appendix}}\end{center}
\begin{center}\large{{\bf Kinematics of inclusive reactions}}\end{center}
For the inclusive reaction  $a+b\rightarrow c+X$, the momentum variables
for the a,b,c particles are defined as follows \parn
$p_a=(E_a, 0, 0, p)$\hspace{1cm} $p_b=(E_b, 0, 0, -p)$ 
\hspace{1cm}$p_c=(E, \vec{p}_T, p_z)$ \parn
The Feynman-x variable is  $x_F\approx \frac{2p_z}{\sqrt s}$.\par
In the large s kinematic limit (where $m_a$ and $m_b$ can be put to zero and  
$m_c^2/s$ neglected), the transformations between  the $(x_F,p_T^2)$ and 
$(t,M_X^2)$ variables  are 
\be \left \{
\begin{array}{l}  
x_F(t,M_X^2)= 1-\frac{M_X^2}{s}+\frac{2t}{s}\\
p_T^2 (t,M_X^2)=-(t-m_c^2)\:(1-\frac{M_X^2}{s}+\frac{t}{s})
\end{array} \right.
\label{transformations2}
\ee
Through the calculation of the Jacobian factor, the differential 
cross sections in terms of two sets of kinematic variables are related as 
follows 
\be
\frac{d^2\sigma}{d x_F\:d p_T^2}=\left (1-\frac{M_X^2}{s}
\right )\;
\frac{d^2\sigma}{dt\:d M_X^2}
\label{cs}
\ee
Using the previous eqs.\ref{transformations2} and \ref{cs}
(where now  a=$\pi^-$, b=$N$, c=$D$), the differential cross section, 
eq.\ref{fitting}, for the $\pi^-\:N\rightarrow D\:X$ scattering in the 
($M_X^2,t$) variables is found to be 
\be
\left[\frac{d^2\sigma}{dM_X^2\:dt}\right]_{EXP}=
\frac{(1-\frac{M_x^2}{s})}{s}\:\left (\frac{M_x^2}{s}-\frac{2t}{s}\right )^n\:
e^{b[m_D^2+(t-m_D^2)\:(1-\frac{M_x^2}{s}+\frac{t}{s})]}
\label{den}
\ee 
Eq.\ref{den} depends on the variable t and in the following we investigate
whether the t/s contribution can be neglected, in the tri-regge
limit. For $m_{\pi}$=$m_N$=0, t is given by 
\be
t(x_F,p_T^2)=m_D^2+x_F\frac{s}{2}\: \left (1-\sqrt{1+\frac{4}{s}\;
\frac{m_D^2+p_T^2}{x_F^2}}\right )
\ee
In Fig.9, we see that t runs from small to rather large values in the 
experimentally  explored region, even for $p_T^2=0$. The  figure refers to 
experiment E769, but the same result applies 
for all  three experiments. 
For the tri-regge limit to be applicable,   t has to be fixed   and 
 $M_X^2/s$  must be small. As a function of 
$x_F$ and $p_T^2$,  $M_X^2/s$ is given by 
\be
\frac{M_X^2}{s}=1+\frac{m_D^2}{s}-x_F\:
\sqrt{1+\frac{4}{s}\;\frac{m_D^2+p_T^2}{x_F^2}}\approx 1-x_F
\ee
so that the small $M_X^2/s$ condition  requires large  $x_F$. 
Returning  to the t variable, we see that t is small at large $x_F$ and 
hence $t/s$ in eq.\ref{den} can be neglected.\par 
Even though data are available  up to 
$x_F^{max}=0.7\div 0.9$, but they are rather imprecise near the boundary, 
we limited our analysis to the reliable region $x_F\approx 0.4\div 0.7$  
($M_X^2/s\approx 0.3\div 0.6$) for testing our predictions.\par
\vskip1mm 
\begin{figure}[htb]
\begin{center}
\epsfig{file=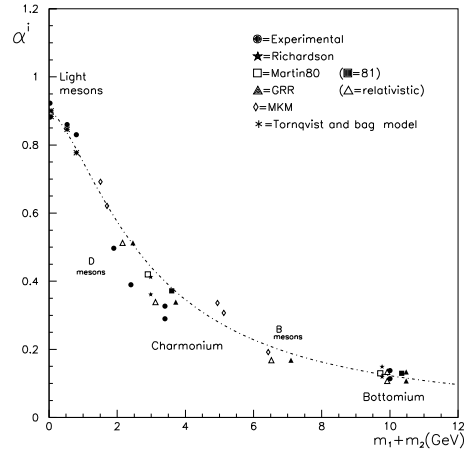,height=6.8cm,width=6.8cm}
\caption{{\em The slope parameter for various mesons 
as a function of $m_1+m_2$.}}
\end{center}
\end{figure}
\vskip1mm   
\begin{figure}[hbt]
\begin{center}
\epsfig{file=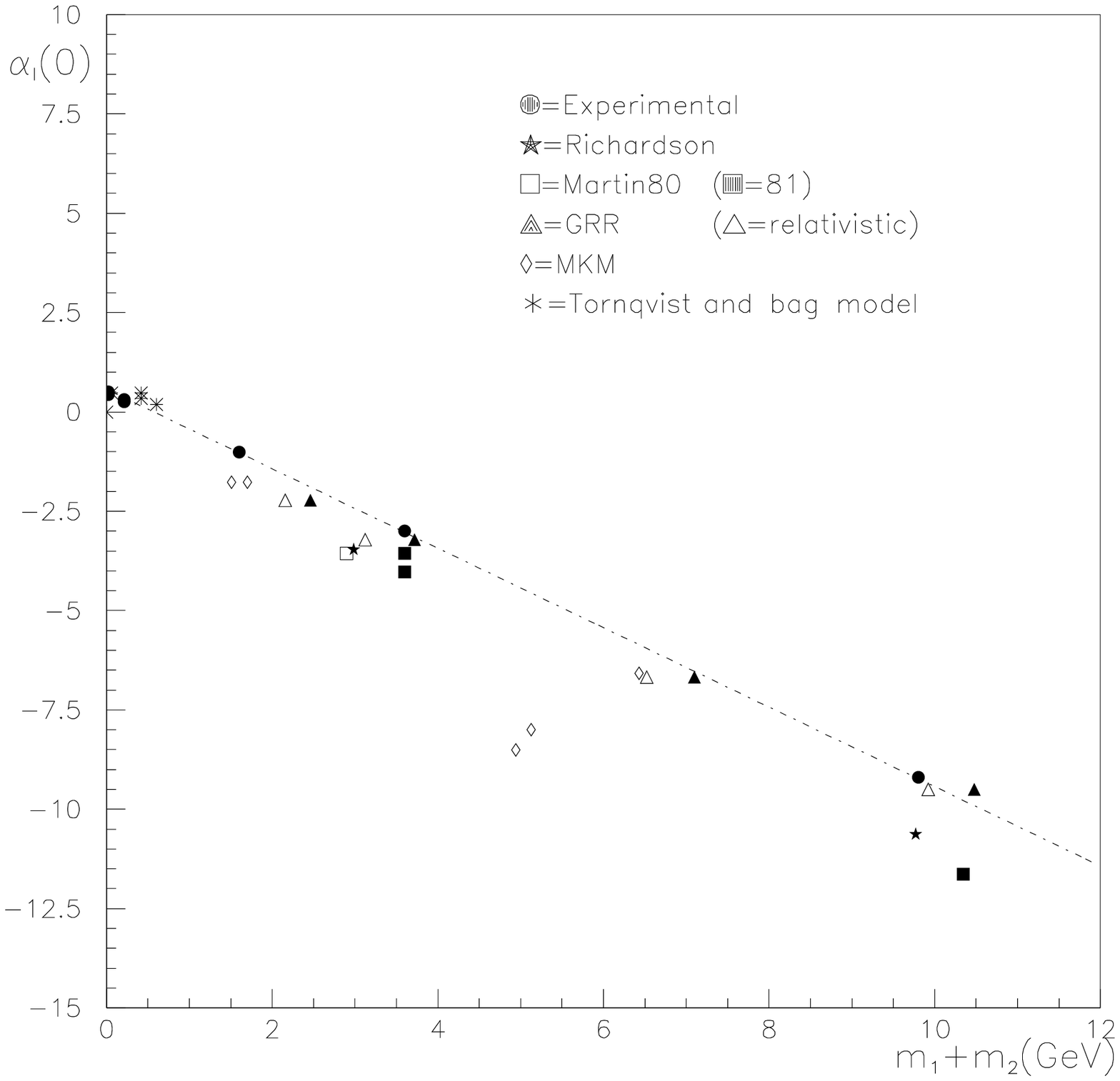,height=6.8cm,width=6.8cm}
\caption{{\em The $\alpha _I(0)$ parameter for various mesons as a function
of $m_1+m_2$.}}
\end{center}
\end{figure}
\vskip1mm 
\begin{figure}[bht]
\epsfig{file=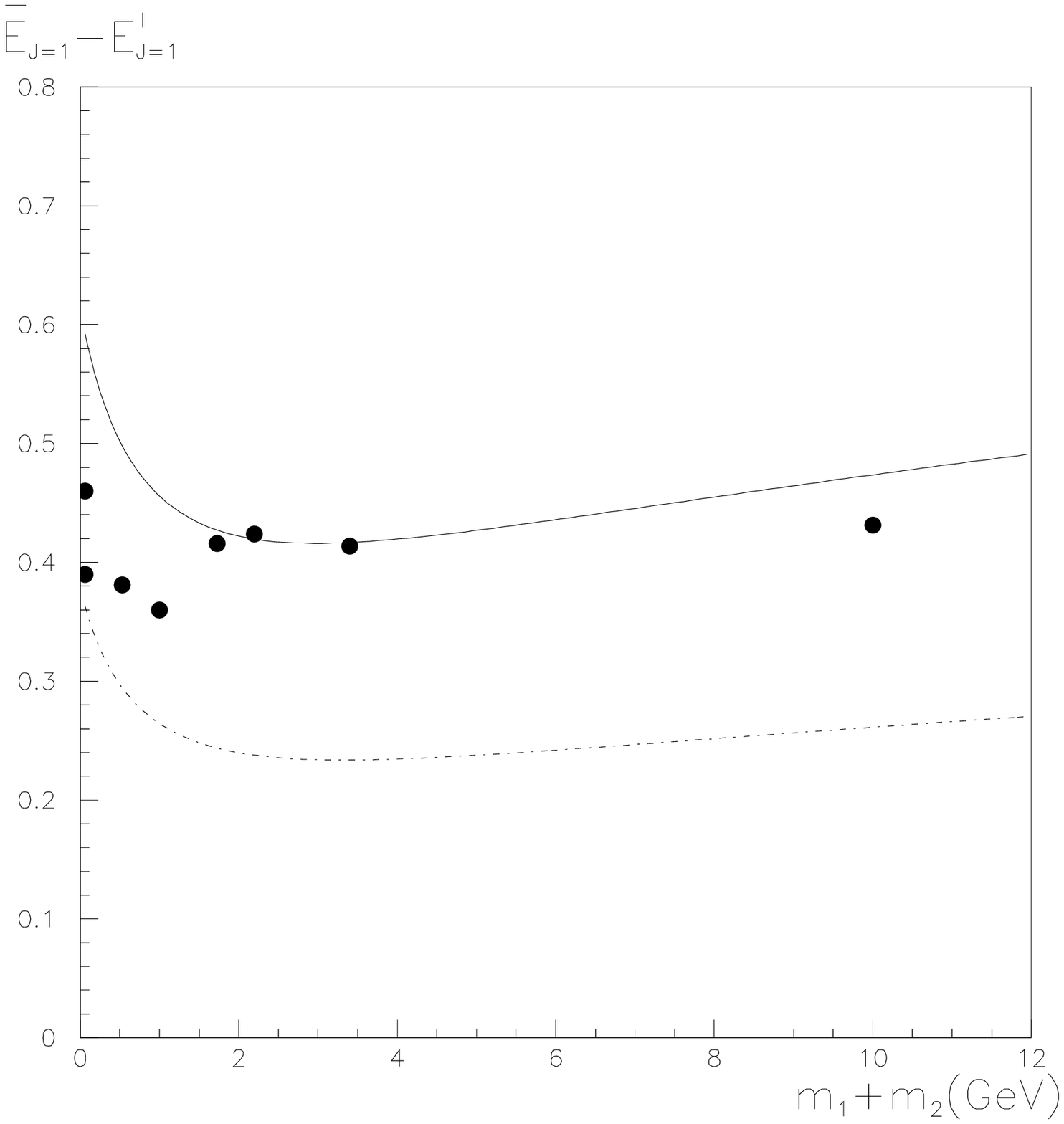,height=6.cm,width=5.8cm}
\epsfig{file=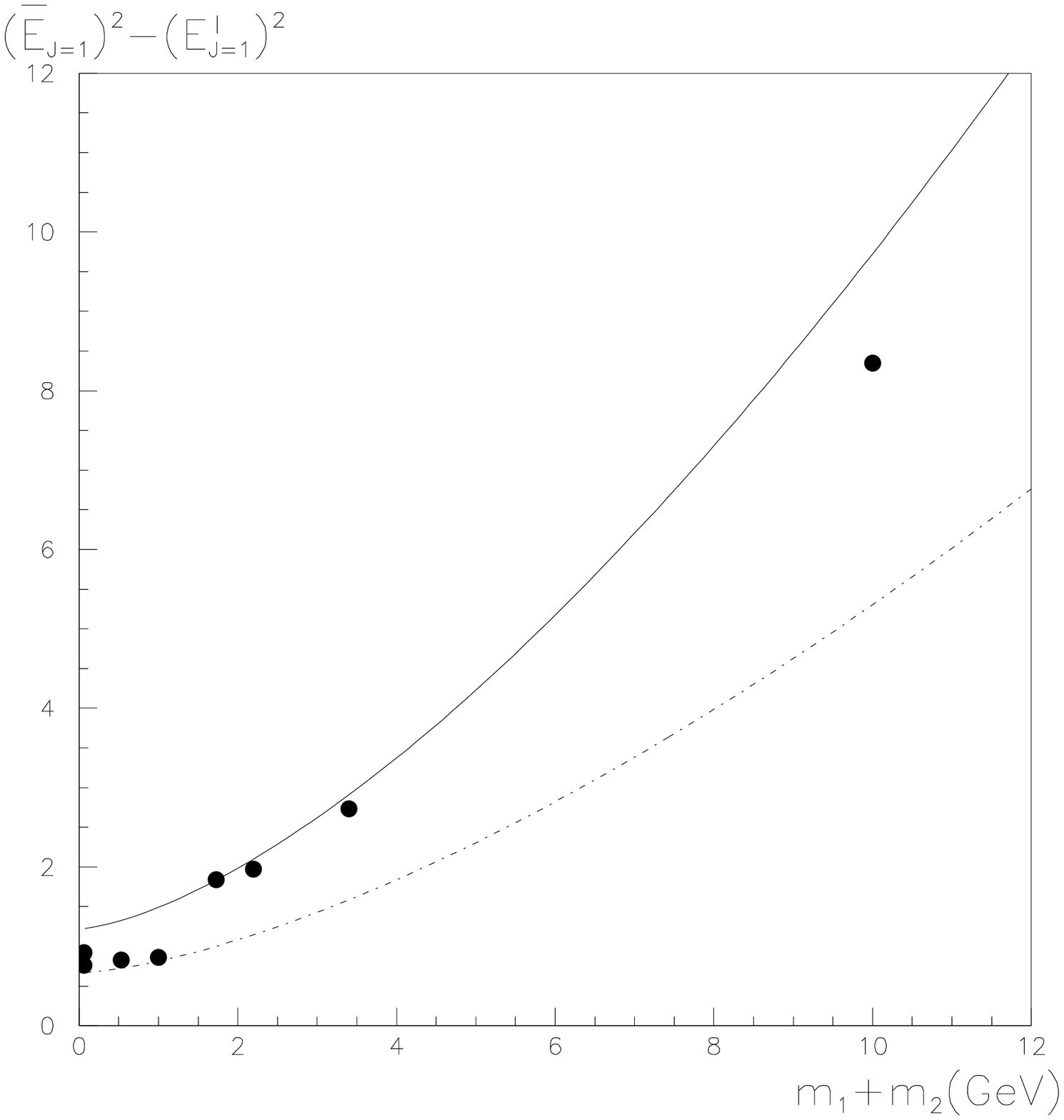,height=6.cm,width=5.8cm}
\caption{Left: {\em The energy splitting  between  the 
 $J^P=1^-$ and $J^P=1^+$ states of various mesons (in Table I) as a
 function of $m_1+m_2$}. Right: {\em Their energy squared gap.
 The curves are for $\bar{d}\alpha (0)=\alpha_I(0)
-\bar{\alpha}(0)=0.6$ [down] and $\bar{d}\alpha (0)=1.1$[up].}}
\end{figure}
\vskip1mm 
\begin{figure}[htb]
\begin{center}
\epsfig{file=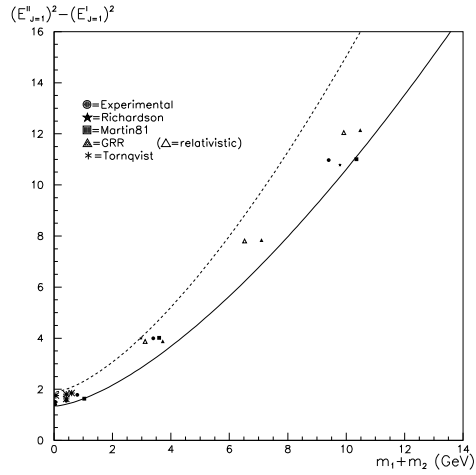,height=6.8cm,width=6.8cm}
\caption{{\em The energy squared  gap between the J=1 states of the 
leading and second trajectory  as a function of $m_1+m_2$. 
The  curves are for $d\alpha (0)=\alpha _{I}(0)-\alpha _{II}(0)$=1.3 [down], 
and $d\alpha (0)$=1.6[up].}}
\end{center}
\end{figure}
\vskip1mm \noindent
\begin{figure}[bht]
\begin{center}
\epsfig{file=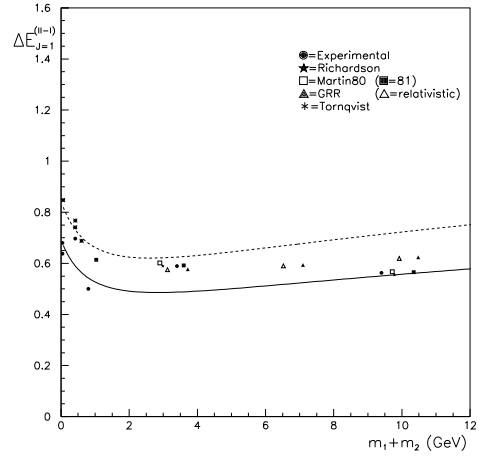,height=6.8cm,width=6.8cm}
\caption{{\em The energy splitting between the J=1 states of the leading 
and second  trajectories as a function of $m_1+m_2$. 
The two curves  are  for $d\alpha (0)=\alpha _{I}(0)-
\alpha _{II}(0)=1.3$ [down] and 
$d\alpha (0)=1.6$ [up].}}
\end{center}
\end{figure}
\vskip1mm \noindent
\begin{figure}[htb]
\begin{center}
\epsfig{file=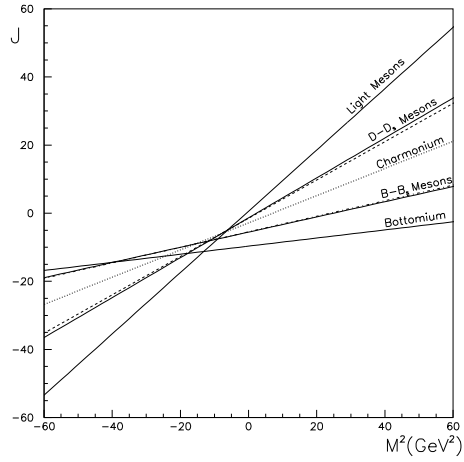,height=6.8cm,width=6.8cm}
\caption{{\em The leading Regge trajectories as derived by our model
(see Table II) in a compared plot.}}
\end{center}
\end{figure}
\vskip1mm
\vskip1mm 
\begin{figure}[htb]
\begin{center}
\epsfig{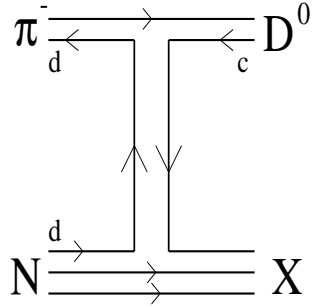}
\end{center}
\caption{{\em A D mesonic trajectory is exchanged in the 
$\pi ^-\:N\rightarrow D\:X$ inclusive reaction.}}
\end{figure}
\vskip1mm\noindent 
\vskip1mm
\begin{figure}[thb]
\begin{center}
\epsfig{file=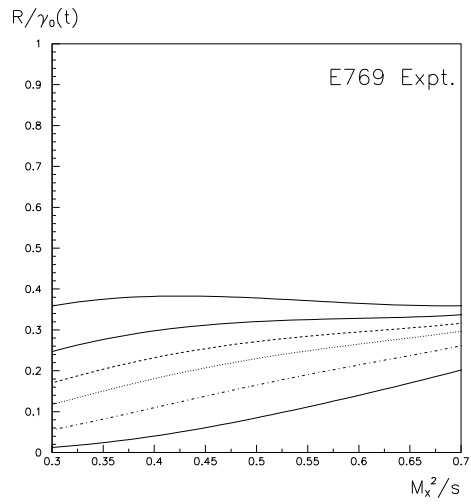,height=6.8cm,width=6.8cm}
\end{center}
\caption{{\em The $R/\gamma_0(t)$ function for fixed and small $t$ values. 
The n and b parameters are from  E769 Expt.}}
\end{figure}
\vskip1mm \noindent
\begin{figure}[thb]
\begin{center}
\epsfig{file=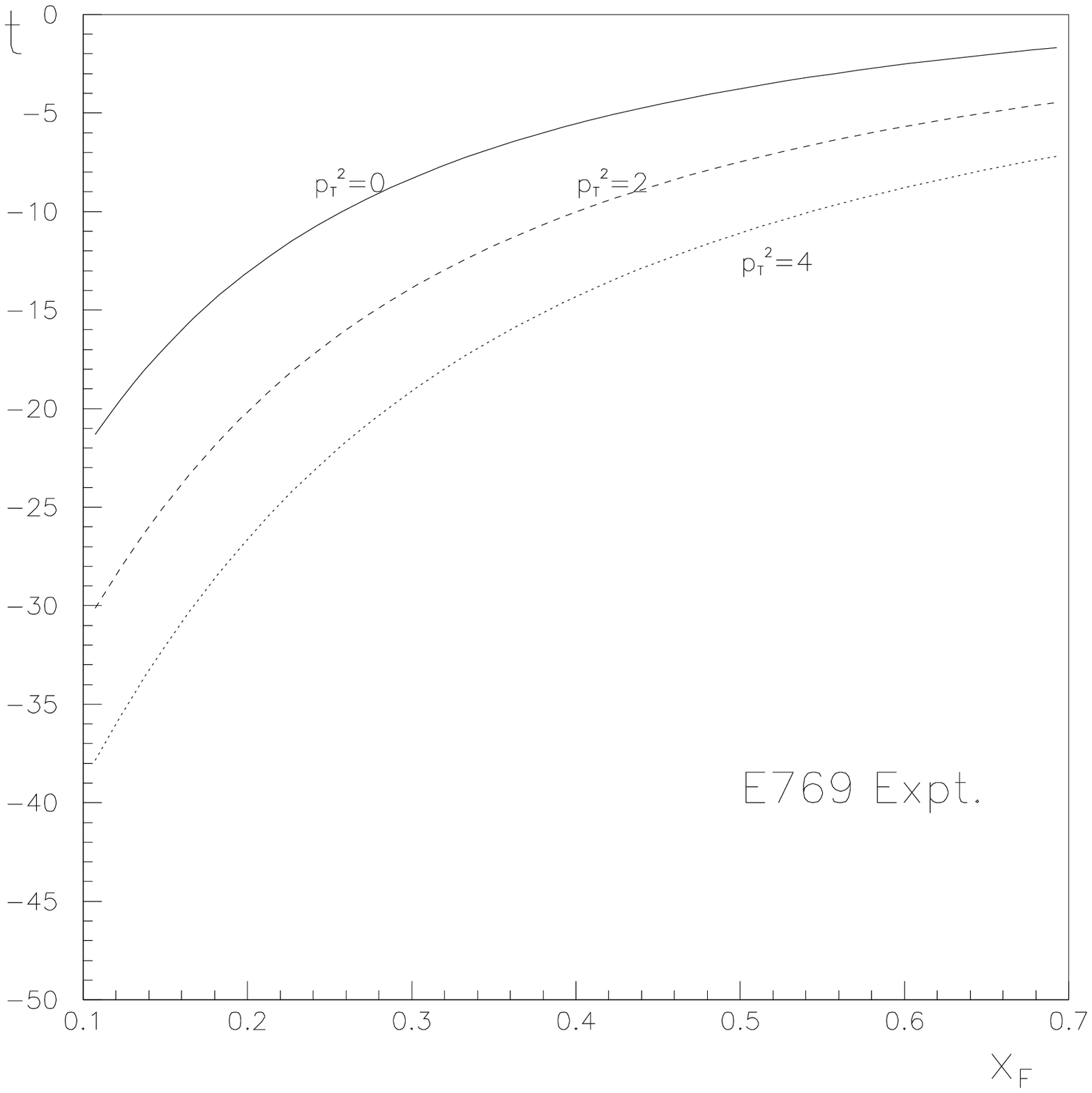,height=6.8cm,width=6.8cm}
\end{center}
\caption{{\em The $t(x_F)$ function for fixed $p_T^2$ values.  
Data are from the E769 Experiment.}}
\end{figure}
\vskip1mm \noindent

\begin{thebibliography}{999}
\bibitem{IMZ91} F. Iachello, N.C.Mukhopadyay and L. Zhang Phys.Rev.{\bf
D}44 (1991), 898. 
\bibitem{DP96} Review of Particle Properties, Phys. Rev. D{\bf 54}
(1996),1-720.
\bibitem{T90} N.A.Tornqvist, in Stockholm 1990, Proceedings, 
{\it Low energy antiproton physics},p. 287-303 and Helsinky University 
preprint HU-TFT-90-52. 
\bibitem{JT76} K.Johnson and C.B.Thorn, Phys. Rev.D{\bf 13} (1976), 1934;
I.Bars and A.J.Hanson, {\em ibid.} {\bf 13} (1976), 1744. For a review of
the bag model see D. Flamm and F. Schoberl, {\it Introduction to the quark
model of elementary particles} (Vol. I). Gordon and Breach Publishers(1982).
\bibitem{R79}J. L. Richardson,   Phys. Lett.  B{\bf 82} (1979) 272.
\bibitem{M80}A. Martin,   Phys. Lett.  B {\bf 93} (1980) 338;   
B {\bf100} (1981) 511. 
\bibitem{GRR93}  A. K. Grant, J. L. Rosner and E. Rynes, Phys. Rev. D 
{\bf47} (1993) 1981. 
\bibitem{MKM88}  J. Morishita, M. Kawaguchi and T. Morii, Phys. Lett. B{\bf185}
(1987) 159;  Phys. Rev. D{\bf 37} (1988) 159.
\bibitem{DA97} DELPHI Collaboration Phys. Lett. {\bf B}398 (1997), 207; 
ALEPH Collaboration Phys.Lett. {\bf B}402, 213. 
\bibitem{KB}A.B.Kaidalov, Z. Phys. C{\bf 12}(1982),63;L.Burakovsky,
L.P.Horwitz and T.Goldman hep-ph/9708468. 
\bibitem{FS97} S.Filipponi and Y. Srivastava, to be published; S.Filipponi,
G.Pancheri and Y. Srivastava, Invited Talk at 
XXVII Symposium on Multiparticle Dynamics, (September 1997, LNF Frascati
Rome), to be published. 
\bibitem{M70} A. H. Muller, Phys. Rev. D{\bf 2} (1970), 2963. 
\bibitem{PS71} G. Pancheri and Y. Srivastava, 
Lett. Nuovo Cimento Vol. II (1971) 381. 
\bibitem{A92} Fermilab E769 Collaboration, G. A. Alves  {\it et al.},   
Phys. Rev. Lett. {\bf69} (1992) 3147;  Phys. Rev.  D {\bf 49} (1994) 4317.
\bibitem{Ag85} Na27 LEBC-EHS Collaboration, M. Aguillar Benitez  
{\it et al.},  Phys. Lett. {\bf161 B} (1985), 400.
\bibitem{B91} Na32 ACCMOR Collaboration, S. Berlag  {\it et al.},
Z. Phys. C {\bf49}(1991), 555.
\bibitem{B90} ACCMOR Collaboration, S. Berlag  {\it et al.}  Phys. Lett.
 B {\bf247} (1990) 113.
\bibitem{NPS84}A. Nakamura, G. Pancheri and Y. Srivastava Zeit.Phys.
{\bf C21}(1984), 243.
\bibitem{DDD} J. Dias de Deus and J. Pulido Zeit.Phys. {\bf C9}(1981),
255.
\end{thebibliography}
\end{document}